\documentclass[preprint2]{aastex}

\shorttitle{The transiting exoplanet WASP-7b}
\shortauthors{Hellier et al.}

\begin{document}

\title{WASP-7: The brightest transiting-exoplanet system in the 
Southern hemisphere.}

\author{
Coel Hellier\altaffilmark{1},
D.R. Anderson\altaffilmark{1}, 
M. Gillon\altaffilmark{2}, 
T.A. Lister\altaffilmark{1,3}, 
P.F.L. Maxted\altaffilmark{1}, 
D. Queloz\altaffilmark{2}, 
B. Smalley\altaffilmark{1}, 
A. Triaud\altaffilmark{2}, 
R.G. West\altaffilmark{4},
D.M. Wilson\altaffilmark{1}, 
K. Alsubai\altaffilmark{5},
S.J. Bentley\altaffilmark{1}, 
A. Collier Cameron\altaffilmark{5}, 
L. Hebb\altaffilmark{5},
K. Horne\altaffilmark{5},
J. Irwin\altaffilmark{6}, 
S.R. Kane\altaffilmark{7}, 
M. Mayor\altaffilmark{2}, 
F. Pepe\altaffilmark{2}, 
D. Pollacco\altaffilmark{8}, 
I. Skillen\altaffilmark{9}, 
S. Udry\altaffilmark{2}, 
P.J. Wheatley\altaffilmark{10},
D.J. Christian\altaffilmark{8}, 
R. Enoch\altaffilmark{11,5}, 
C.A. Haswell\altaffilmark{11}, 
Y.C. Joshi\altaffilmark{8}, 
A.J. Norton\altaffilmark{11}, 
N. Parley\altaffilmark{11}, 
R. Ryans\altaffilmark{8}, 
R.A. Street\altaffilmark{3,8}, 
I. Todd\altaffilmark{8} 
}

\altaffiltext{1}{Astrophysics Group, Keele University, Staffordshire, ST5 5BG, UK}
\altaffiltext{2}{Observatoire de Gen\`{e}ve, 51 ch. des Maillettes, 1290 Sauverny, Switzerland}
\altaffiltext{3}{Las Cumbres Observatory, 6740 Cortona Dr. Suite 102, Santa Barbara, CA 93117, USA}
\altaffiltext{4}{Department of Physics and Astronomy, University of Leicester, Leicester, LE1 7RH, UK}
\altaffiltext{5}{School of Physics and Astronomy, University of St. Andrews, North Haugh,  Fife, KY16 9SS, UK}
\altaffiltext{6}{Department of Astronomy, Harvard University, 60 Garden
Street, MS 10, Cambridge, Massachusetts 02138, USA}
\altaffiltext{7}{Michelson Science Center, Caltech, MS 100-22, 770 South Wilson Avenue, Pasadena, CA 91125, USA}
\altaffiltext{8}{Astrophysics Research Centre, School of Mathematics \& Physics, Queen's University, University Road, Belfast, BT7 1NN, UK}
\altaffiltext{9}{Isaac Newton Group of Telescopes, Apartado de Correos 321, E-38700 Santa Cruz de la Palma, Tenerife, Spain}
\altaffiltext{10}{Department of Physics, University of Warwick, Coventry, CV4 7AL, UK}
\altaffiltext{11}{Department of Physics and Astronomy, The Open University, Milton Keynes, MK7 6AA, UK}

\begin{abstract}
We report that a Jupiter-mass planet, WASP-7b, transits the $V$ = 9.5
star HD\,197286 every 4.95 d.  This is the brightest discovery from
the WASP-South transit survey and the brightest transiting-exoplanet
system in the Southern hemisphere.  WASP-7b is among the densest of 
the known Jupiter-mass planets, suggesting that it has a massive core. 
The planet mass is 0.96$^{+0.12}_{-0.18}$
M$_{\rm Jup}$, the radius 0.915$^{+0.046}_{-0.040}$ R$_{\rm Jup}$
and the density 1.26$^{+0.25}_{-0.21}$ $\rho_{\rm Jup}$.  
\end{abstract}


\section{Introduction}
Transiting exoplanets are valuable discoveries since they offer the most
opportunities for parametrization and study.  The WASP project
\citep{sw} is one of a number of wide-area surveys, along
with HAT \citep{hat}, TrES \citep{tres} and XO \citep{xo}, all
aimed at finding exoplanets transiting relatively bright stars, where
they are easiest to observe.  A prime aim is to fill out diagrams 
such as the exoplanet mass--radius plot, which has the potential to be
a diagnostic tool for exoplanets comparable to the Hertzsprung--Russell 
diagram for stars.

WASP is the only one of the above surveys operating in both
hemispheres. We report here on WASP-7 (= HD\,197286), a new discovery
from WASP-South that, at magnitude 9.5, is the brightest
transiting-exoplanet system in the Southern hemisphere, being three
magnitudes brighter than the brightest previously announced, WASP-4
\citep{wasp4} and WASP-5 \citep{wasp5}.  WASP-7b is also among the
densest known Jupiter-mass exoplanets, extending the populated region
of the mass--radius plot.

\begin{figure}
\hspace*{-5mm}\includegraphics[width=9cm]{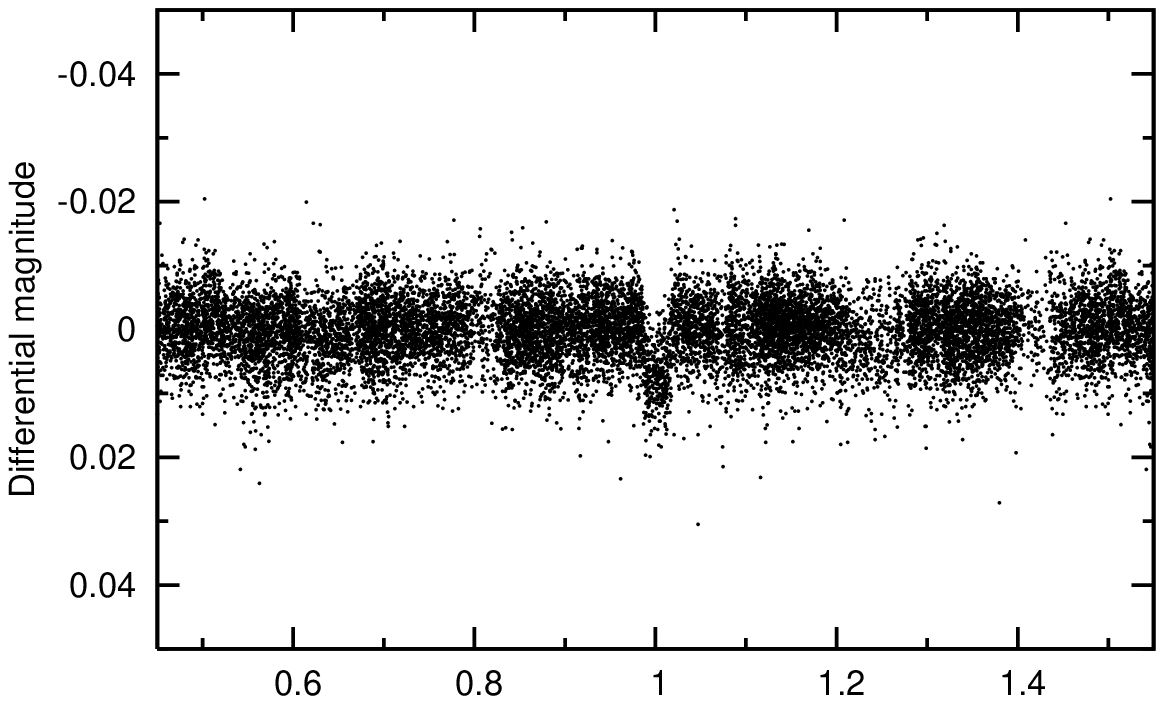}\\
\hspace*{-5mm}\includegraphics[width=9cm]{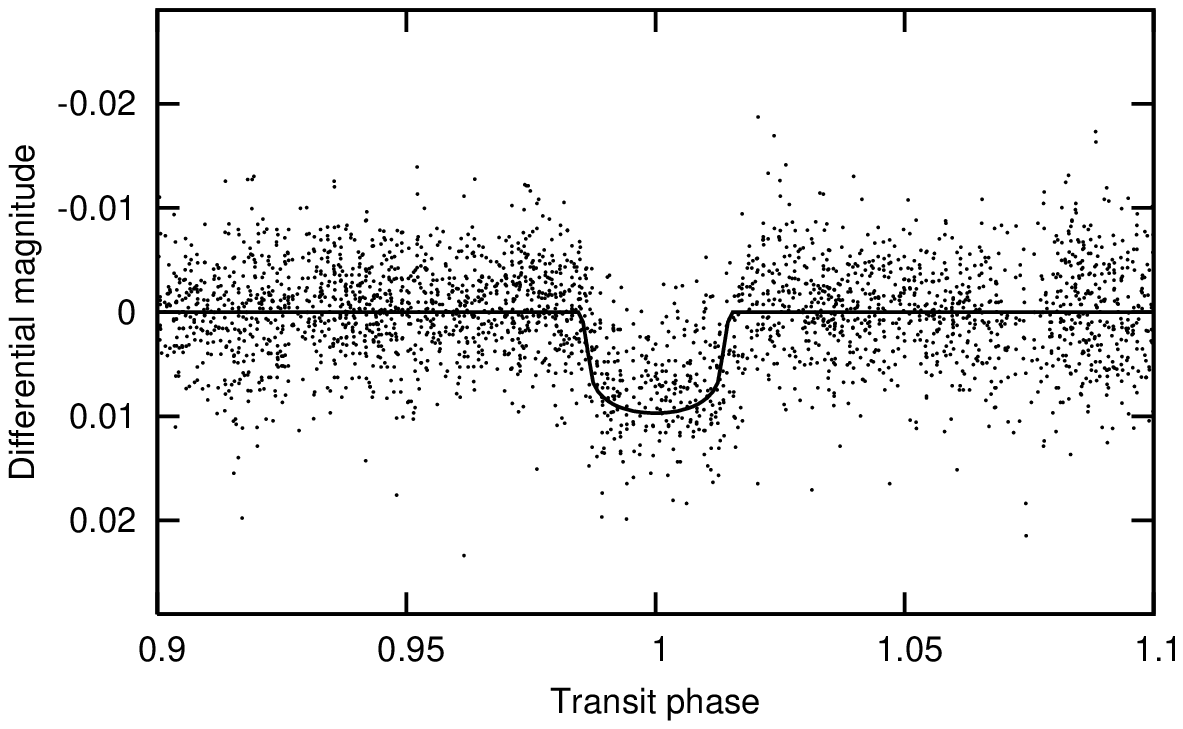}
\caption{The WASP-South lightcurve of WASP-7 folded on the 4.95-d
transit period. The lower panel shows the transit phases enlarged,
together with the best-fitting model.}
\label{figure:wasp}
\end{figure}

\begin{table}
\caption{CORALIE radial velocities for WASP-7.\protect\rule[-1mm]{0mm}{4mm}}
\begin{tabular}{cccc} \tableline\tableline Time \rule{0mm}{4mm}& Rad\,Vel &
 $\sigma _{RV}$ & Bis\,Span \\ 
BJD\,$-$\,245\,0000 & (km\,s$^{-1}$) & (km\,s$^{-1}$) & (km s$^{-1}$)$^{a}$ 
\\ \tableline
4329.71524 & $-$29.8238 &  0.0214 & $-$0.344\rule{0mm}{4mm} \\
4330.79865 & $-$29.7059 &  0.0182 & $-$0.558  \\ 
4359.54361 & $-$29.8400 &  0.0196 & $-$0.703 \\ 
4362.52524 & $-$29.9620 &  0.0202 & $-$0.529  \\
4364.55740 & $-$29.7575 &  0.0221 & $-$0.691  \\
4376.57881 & $-$29.8540 &  0.0171 & $-$0.669  \\
4378.59076 & $-$29.8662 &  0.0194 & $-$0.736  \\
4379.57799 & $-$29.7858 &  0.0222 & $-$0.547  \\
4380.63008 & $-$29.8169 &  0.0173 & $-$0.571  \\
4382.66066 & $-$29.9355 &  0.0210 & $-$0.641  \\
4386.61194 & $-$29.8845 &  0.0229 & $-$0.673  \\ \tableline
\end{tabular} \\$^{a}$ Bisector spans; 
$\sigma$$_{BS}$ $\approx$ 2\,$\sigma$$_{RV}$\rule{0mm}{4mm}
\end{table}

\begin{figure}
\hspace*{-5mm}\includegraphics[width=9cm]{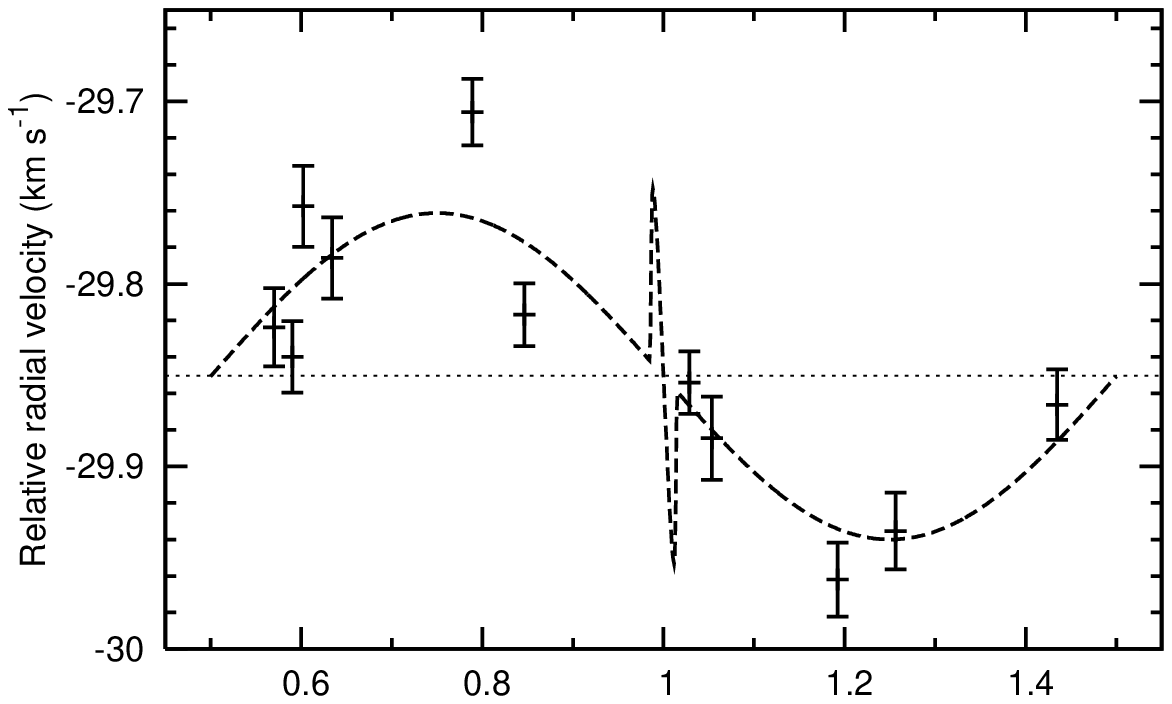}\\
\hspace*{-5mm}\includegraphics[width=9cm]{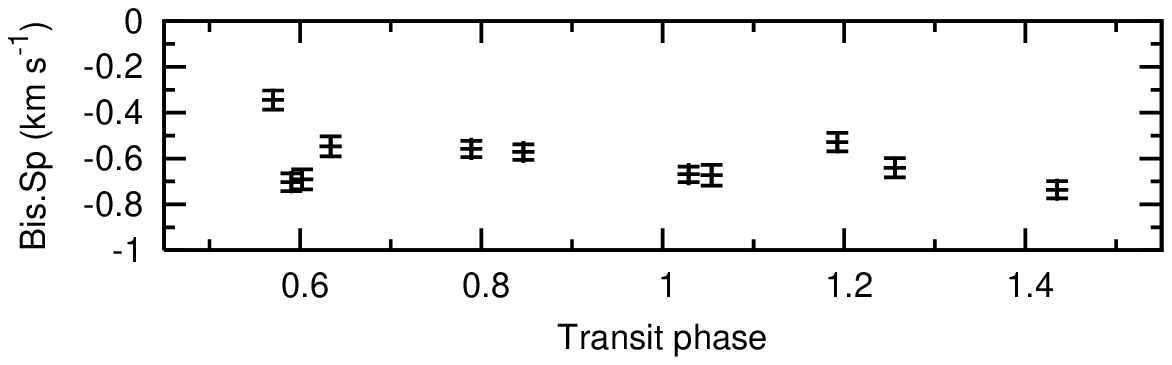}\\
\caption{The CORALIE  radial-velocity curve for WASP-7 along
with the best-fitting model (including the predicted 
Rossiter--McLaughlin effect,
which is large for this fast rotator).  The lower panel shows the 
bisector spans. }
\end{figure}

\section{Observations}
The WASP-South survey is described in \cite{sw}
and \cite{wasp4}, while a discussion of our candidate
selection methods can be found in \cite{wasp1}, 
\cite{wasp3}, and references therein. 

WASP-7 (= HD\,197286) is a $V$ = 9.5 F5V star in Microscopium.  
It was observed by WASP-South
from May to mid-October in both 2006 and 2007, being recorded 
in two overlapping cameras, each an  11.1-cm-aperture
Canon 200-mm f/1.8 lens backed by a 2k$\times$2k $e$2$V$ CCD.
Exposure times were 30 s, with a typical cadence of 8 minutes.  
We obtained 5800 photometric data points from each
camera in 2006, and a further 5700 data points from each 
in 2007.  

The WASP-South lightcurves revealed dips with a depth of
0.007 magnitudes recurring with a 4.95-d period (Fig.~1). 
Spectroscopic observations were then obtained using the
CORALIE spectrograph on the Euler 1.2-m telescope. 
Eleven radial-velocity measurements were obtained during
2007 Sept 15 to Oct 12 (Fig.~2; Table~1), 
establishing WASP-7b as a Jupiter-mass companion.

We used a line-bisector analysis to look for asymmetries in the
spectral line profiles, as could be caused by contamination from an
unresolved eclipsing binary \citep{bisector}. 
Such a binary would produce bisector spans that
vary in phase with the photometric period with an amplitude comparable
to the radial-velocity amplitude. This is not seen in our data, 
supporting the conclusion that the radial-velocity variations are caused 
by a planet.

\begin{table}
\caption{Parameters of WASP-7/HD\,197286.\protect\rule[-1mm]{0mm}{4mm}} 
\begin{tabular}{lrl}
\tableline\tableline 
\multicolumn{3}{l}{R.A. = 20$\rm^{h}$44$\rm^{m}$10.22$\rm^{s}$, Dec. =
--39$^\circ$13$^{'}$30.9$^{''}$\rule{0mm}{5mm}} \\ [2mm]
$T_{\rm eff}$  & 6400 $\pm$ 100 & K \\ 
log ${\it g}$ & 4.3 $\pm$ 0.2 \\ 
$[$Fe/H$]$ & 0.0 $\pm$ 0.1 \\
$\xi_{\rm t}$ & 1.5 $\pm$ 0.2 & km\,s$^{-1}$ \\
$v\sin i$     & 17 $\pm$ 2 & km\,s$^{-1}$ \\
Spectral Type & F5V \\
Distance &  140 $\pm$ 15 & pc \\ 
 $V$ mag & 9.51 \\ \tableline
\end{tabular}
\label{wasp7-params}
\end{table}

\section{WASP-7 (HD\,197286) parameters}
The CORALIE spectra, when co-added, give a
signal-to-noise ratio of $\sim$\,50 in 0.01\,\AA\ bins, which is
suitable for a preliminary photospheric analysis of WASP-7. 

We analysed the spectra using the {\sc uclsyn} 
package and {\sc atlas9} models, without convective overshooting
\citep{castelli}, leading to the parameters
in Table~2.  The  effective temperature ($T_{\rm eff}$) comes from 
an analysis of the H$\alpha$ line while the  surface gravity 
($\log g$) comes from
the  Na\,{\sc i} D and Mg\,{\sc i} b lines.  An estimate of
the microturbulence ($\xi_{\rm t}$) comes from  several clean and unblended
Fe\,{\sc i} and Fe\,{\sc ii} lines, while the  ionization balance
between Fe\,{\sc i} and Fe\,{\sc ii} was used as an additional 
diagnostic of $T_{\rm eff}$ and $\log g$. 

In addition to the spectral analysis, we have also used TYCHO, DENIS
and 2MASS magnitudes to estimate the effective temperature using the
Infrared Flux Method \citep{blackwell}. This gives $T_{\rm eff}$  
= 6370 $\pm$ 150~K, which is in close agreement with that obtained from
the spectroscopic analysis. These results are consistent with the F5V
spectral type determined by \cite{houk}. 

The  rotation rate of WASP-7 is $v \sin i$ = 17\,$\pm$\,2 km\,s$^{-1}$.
This is in line with expectations for an F5V star (Gray 1992 quotes
$<$\,$v \sin i\,$$>$ = 20), but is the largest
known among host stars of transiting planets, and leads to the
prediction of a large Rossiter--McLaughlin effect of 98\,$\pm$\,19 m
s$^{-1}$ (e.g.\ Gaudi \&\ Winn 2007).  

We have investigated whether the star might be chromospherically active, 
however there is no sign of Ca H+K line emission in the CORALIE
spectra.  We have also looked for variability in the
WASP data at the predicted rotation period of 3.7\,$\pm$\,0.5 d,
but find no such variability with an upper limit of 0.02 mag.  

The Li {\sc i} 6708\AA\ line is not detected in the 
CORALIE spectra (EW $\ll$ 1m\AA, allowing us to derive an upper-limit on
the lithium abundance of log n(Li/H) + 12 $\ll$ 1.0). However, the
$T_{\rm eff}$ of this star implies that it is in the lithium gap
\citep{lithgap} and thus the lithium line does not provide an
age constraint.

\begin{table}
 \caption{System parameters for WASP-7.\protect\rule[-1mm]{0mm}{4mm}}
 \label{sys-params}
 \begin{tabular}{lcc}
 \hline
 Parameter & \multicolumn{2}{c}{Value}\rule{0mm}{4mm}\\
 \hline
 $P$ (days)\rule{0mm}{4mm} & 4.954658 & $^{+ 0.000055}_{- 0.000043}$\\
 $T_{\rm C}$ (HJD) & 2453985.0149 & $^{+ 0.0009}_{- 0.0012}$\\
 $T_{\rm 14}$ (days)$^{a}$ & 0.1573 & $^{+ 0.0024}_{- 0.0018}$\\
 $R_{\rm P}^{2}$/R$_{*}^{2}$ & 0.00579 & $^{+ 0.00013}_{- 0.00026}$\\
 $b$ $\equiv$ $a \cos i/R_{\rm *}$ (R$_{\odot}$) & 0.08 & $^{+0.17}_{-0.08}$\\
 $i$ (degs) & 89.6 & $^{+0.4}_{-0.9}$\\
\medskip
 $e$ & 0 (adopted)\\
 $K_{\rm 1}$ (km s$^{-1}$) & 0.097 & $^{+ 0.013}_{- 0.013}$\\
 $\gamma$ (km s$^{-1}$) & $-$29.8506 & $^{+ 0.0017}_{- 0.0016}$\\
 $M_{\rm *}$ (M$_{\odot}$) & 1.28 & $^{+ 0.09}_{- 0.19}$\\
 $R_{\rm *}$ (R$_{\odot}$) & 1.236 & $^{+ 0.059}_{- 0.046}$\\
 $T_{\rm eff}$ (K) & 6400 & $\pm$\,100\\
\medskip
 $\log g_{*}$ (cgs) & 4.363 & $^{+ 0.010}_{- 0.047}$\\
 $M_{\rm P}$ (M$_{\rm J}$) & 0.96 & $^{+ 0.12}_{- 0.18}$\\
 $R_{\rm P}$ (R$_{\rm J}$) & 0.915 & $^{+ 0.046}_{- 0.040}$\\
 $\rho_{\rm P}$ ($\rho_{\rm J}$) & 1.26 & $^{+ 0.25}_{- 0.21}$\\
 $a$ (AU) & 0.0618 & $^{+ 0.0014}_{- 0.0033}$\\
 $\log g_{\rm P}$ (cgs) & 3.421 & $^{+ 0.067}_{- 0.071}$\\
 $T_{\rm P}$ (K) & 1379 & $^{+ 35}_{- 23}$\\ [1mm] 
 \hline
\multicolumn{3}{l}{$^{a}$ $T_{\rm 14}$: duration, time from 1$^{\rm st}$ to 
4$^{\rm th}$ contact.\rule{0mm}{4mm}}\\
\multicolumn{3}{l}{Errors are 1-$\sigma$}
 \end{tabular}
\end{table}

\begin{figure}
\hspace*{-5mm}\includegraphics[width=9cm]{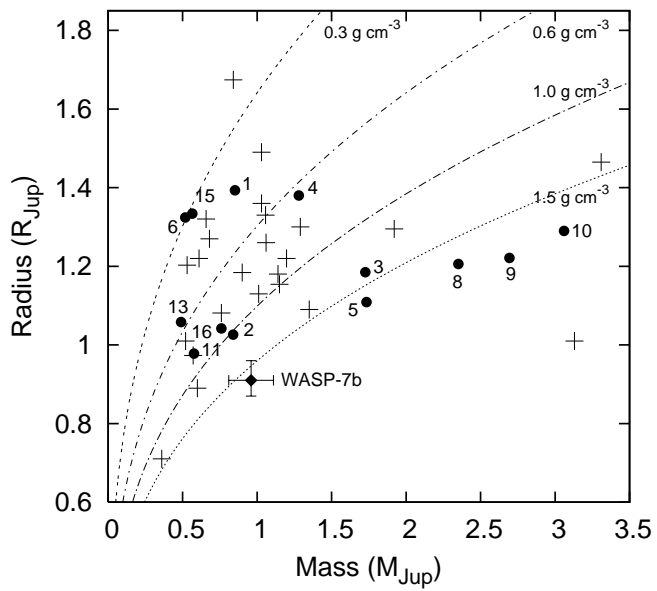}
\caption{The mass--radius diagram for Jupiter-mass transiting 
exoplanets. The WASP planets (points) are labelled by their number
(data from the WASP consortium, papers in preparation).   
The remaining data (crosses) are taken from \cite{pont}.}
\end{figure}

\subsection{Planetary parameters}
The CORALIE radial-velocity measurements were combined with the
WASP-South photometry in a simultaneous Markov-chain Monte-Carlo
(MCMC) analysis to find the parameters of the WASP-7 system.
This process is described in detail in \citet{mcmc}
and \citet{wasp3}.   The optimal MCMC
parameters are listed in Table~\ref{sys-params}.

An initial run found an eccentricity, $e < 0.17$, and we then adopted
 $e =0$ for the solution presented here.  In order to balance the 
weights of the photometry and radial velocities in the MCMC analysis 
we added a systematic error of  20 m s$^{-1}$ to the radial
velocities (as might arise, for example, from stellar activity) to reconcile
$\chi^{2}$ with the number of degrees of freedom. 

WASP-7b is among the densest of the Jupiter-mass planets (Fig.~3), 
which suggests that it has a large core.  The calculations presented
by Fortney, Marley \&\ Barnes (2007) suggest a core mass of $\approx$\,100
Earth masses, or $\approx$\,0.3 of the planet, with some dependence 
on the planetary age. 
Overall, WASP-7b adds to the finding that Hot Jupiters show a 
wide disparity in densities, from the denser WASP-5b \citep{wasp5}
and WASP-7b to the bloated TrES-4 (Mandushev et\,al.\ 2007) 
and WASP-12b (Hebb et\,al.\ 2008).

\acknowledgments
\section{Acknowledgments}
The WASP consortium comprises the Universities of Keele, Leicester,
St.~Andrews, the Queen's University Belfast, the Open University and
the Isaac Newton Group.  WASP-South is hosted by the South African
Astronomical Observatory and we are grateful for their support
and assistance. Funding for WASP comes from consortium universities
and from the UK's Science and Technology Facilities Council.

\end{document}